# Smart Speakers, the Next Frontier in Computational Health

*Jacob Sunshine, MD* [1,2]

**Affiliations:**
[1] University of Washington School of Medicine, Department of Anesthesiology & Pain Medicine, University of Washington, Seattle, WA, USA
[2] Paul G. Allen School of Computer Science and Engineering, University of Washington, Seattle, WA, USA

**Contact:** jesun@uw.edu

## Introduction

The rapid dissemination and adoption of smart speakers has enabled substantial opportunities to improve human health. Just as the introduction of the mobile phone led to considerable health innovation, or ultrasound enabled new opportunities for point of care diagnosis and procedural optimization, smart speaker computing systems carry several unique advantages that can catalyze new fields of research, particularly in out-of-hospital environments. The recent rise and ubiquity of these smart computing systems, which are often cheaper than smartphones and substantially less expensive than medical-grade equipment, hold significant potential for enhancing chronic disease management, enabling passive identification of unwitnessed medical emergencies, detecting subtle changes in human behavior and cognition, limiting isolation, and potentially allowing widespread, passive, remote monitoring of respiratory-based infectious diseases which impact the public health—all while still providing general utility for users. Advances in machine-based classification of disease states, capable of being run on-device and securely in the cloud, can enable rapid diagnostic and predictive functions at low cost, while preserving privacy. This confluence of factors has created a significant opportunity around these devices, which currently reside in 1 of 4 US households, when applied thoughtfully to carefully chosen health conditions.[1]

## What are smart speakers, what are the key differences between smartphones?

At its most basic, a smart speaker is a system comprising a speaker, a microphone array, an embedded computer, a software and machine learning-based intelligent assistant, and wireless connectivity that enables data integration with the cloud, nearby smart devices and other IT infrastructures outside of the home. The increasing computational horsepower of embedded platforms coupled with advances in machine learning have enabled on-device capabilities that removes the need to transmit audio to the cloud.  As such, the system has the capability to continuously monitor the home environment and instruct or converse with a patient about a medically-relevant topic, identify health-related audible biomarkers, sense the environment for contextually relevant health-related motion and much more. And because these computing systems have wireless capability, they can transmit data to the cloud for secure storage and analysis, if desired.  Such connectivity also, in theory, enables integration with medical IT infrastructures so a trained provider can interpret, triage and act upon relevant information from a smart speaker or, in an emergent context, connect with an emergency response system (e.g., 911) to summon help. Key differentiators of these devices compared to mobile phones include that they are plugged in, thus avoiding power constraints that are associated with charging a device; they are predominantly stationary, enabling long-term, passive and continuous monitoring; and their range of measurements is greater than a phone, which generally must be interacted with when it is directly in a user's hands. The inherent constraints of their placement, moreover, provides a substantive benefit by reducing the number of "edge" cases that invariably arise when building intelligent sensing systems. Yet perhaps smart speakers' biggest advantage over mobile phones and other wearable devices is their ability to foster compliance[2,3] by not requiring patients to wear or do anything after initial setup (i.e., they can be truly "set and forget").

Against this background, there are 3 broad mechanisms for how a smart speaker can interact with a patient to improve health. These include (i) as an intelligent conversational agent, (ii) a passive identifier of medically relevant diagnostic sounds and (iii) active sensing using the device's internal hardware to measure physiologic parameters, such as with active sonar, radar or computer vision.

**Smart speakers as health conversation agents**

The first deployed, and most straightforward use for smart speakers are as intelligent conversational agents and facilitators. In the home environment, conversational use cases include the system providing reminders to take medications, retrieving recent lab results (e.g., blood sugar), managing medical appointments and tracking wellness goals.[4] These systems are also capable of reducing isolation, particularly in older adults, by providing a low barrier way to facilitate communication (with family members, caretakers, social workers), including based on detected signals in the environment where a check-in may be warranted (e.g. a change or reduction in activities of daily leaving). Outside of the home, these devices also have a role in the clinic and the inpatient environment. Within the clinic, these devices may soon be used to help liberate physicians from their computers, as provider-patient conversations are passively captured, parsed and analyzed to efficiently document medical encounters.[5] Devices have also been deployed in hospitals, particularly patient rooms, primarily as a way to improve the patient experience[6] and, in the setting of COVID-19, to provide a crucial means of communication with the care team and family unable to visit.[7]

**Classification of medically-relevant diagnostic sounds**

The next level of interaction with these devices is as a classifier of medically-relevant, contextually appropriate bio-signals that represent signs and symptoms of disease. There have been major advances in sound classification research in the computing community[8,9,10] that have crossover for medically-informative audio.[11] Researchers are looking at publicly available datasets from the computing community like Audio Set[12] to relabel and train new models for medically-relevant sounds.[13] In this use case, which would predominate in home environments, the computing system classifies certain audible biomarkers for the purposes of diagnosis or to better inform disease management. Similar to invoking certain trigger words (e.g., "Hey Siri", "Alexa", "Hey Bixby," "OK Google"), these systems are capable of passively identifying specific audio signatures that are contextually relevant and of medical utility. Building on classification guidance from NIH and FDA, Coravos et al. has proposed a useful framework of 'digital biomarkers', which classifies signals as they relate to susceptibility or risk, diagnosis, monitoring, prognostication and prediction.[14] These audio biomarkers can be used to detect and classify cough[15,16], discern voice changes arising from neurodegenerative diseases such as Parkinson's[17] or dementia[18], characterize voice changes related to depression[19,20] or other mental illnesses[21], classify breathing patterns associated with obstructive sleep apnea[22], identify deteriorating asthma[23], and even identify unwitnessed cardiac arrest by detecting the presence of agonal breathing.[11]

**Active sensing using smart devices**

The final way that these computing systems can be used is perhaps the most innovative, and involves turning these devices into contactless active sensing systems using computer vision, sonar or radar for the purposes of physiologic monitoring. If the smart speaker has a camera, this enables important diagnostic capabilities aided by computer vision, which enables a machine to make inferences based on dynamic images and subtle changes in pixilation. Notable potential use cases for computer vision include detection of falls[24], respiratory and heart rate monitoring[25,26], identifying significant changes in activity in older populations[27], self-monitoring of physical therapy, monitoring of acute and chronic wounds[28], and post-operative and post hospitalization-based rehabilitation within the home. In addition, because these devices have speakers and microphones, they are capable of active sonar and echolocation utilizing high (>18 kHz), inaudible frequencies to detect medically relevant motion. Some smart speakers are already enabling these features for activity sensing and gesture detection. A benefit of this method is that, because it utilizes inaudible frequencies, it can collect relevant data while filtering out all audible speech and thus preserves privacy. Similarly, in radar-based systems, electromagnetic waves are transmitted into the environment and phase changes in the reflected signals can be used to classify medically-relevant motion. The potential use cases of these sonar- and radar-based active sensing modalities include monitoring of chest motion/breathing[29] and its perturbations (pertinent for asthma[30], COPD[31], OSA[32], opioid overdose[33]), sleep disturbance (e.g., insomnia), identification of incipient respiratory infection, measurement of cardiac activity, monitoring of activity levels based on movement, epilepsy monitoring and more.

**Privacy**

As with any ubiquitous computing system, a critical consideration relates to privacy, which can mean different things to different people. For a health monitoring context, this refers to monitoring that, similar to default functionality of these devices, enables continuous "listening", but only processes and stores (if the user desires) relevant health data. In practice, using asthma or COPD as an example, the system would not store or analyze conversations, though it would recognize, document and analyze increases in nocturnal cough or relevant changes in respiration such as dyspnea or audible wheeze. It is important that any health-related data approved to be stored is done securely within an environment designed for HIPAA, belong to and be made easily accessible to individuals, and be GPDR-compliant. Just as there are potential privacy concerns associated with smartphones and personal computers, there is a point where their utility outweighs their real and perceived privacy concerns and trust issues. The adoption arc for smart speakers is undoubtedly affected by these concerns, presenting a challenge but also an opportunity to develop innovative privacy-preserving functionality that would make collection of health data more comfortable and trustworthy. These efforts would be greatly enhanced by manufactures taking straight-forward, transparent actions that foster trust and maximize control of information for the monitored user.

**Barriers to implementation/future directions**

While there is tremendous potential for this new computing platform to potentially improve human health, there remain several barriers. The first major barrier is the lack of an open ecosystem, compared to the development environment and regulatory frameworks for applications that can run on smartphones, tablets or PCs. Crucially, there is no "app store" or developer environment that provides the level of access to firmware that would enable flexible development of innovative, high-quality, medically-relevant applications which take advantage of a device's internal hardware. For example, unlike on Android or iOS, a developer cannot leverage the smart speaker's camera, individual speaker(s), or microphones for the purposes of app development. While the major smart speaker manufactures allow for development of "Skills" or plug-ins within a highly-constrained design framework, including at least one enabling secure transmission of health information[34], they do not offer the openness and flexibility that exists for development of health-related applications intended for smartphones. Such an ecosystem would represent a substantial opportunity for health-related software development and would leverage these devices' full computational capabilities.

Control of data flow for regulatory and HIPAA standards are also critical in healthcare use cases. Regulatory organizations, health system stakeholders, and computing communities need to come together to develop a shared understanding on the responsible use of data for these emerging technologies. In particular, it is unclear what protections are needed for data generated in the home that *could* be used for health purposes compared to data that is generated in a clinic or hospital encounter, where protections are clearly enumerated for patient data. Current regulatory guidance do not take into account these new sources of data generated in the home and will have to be addressed as these computing systems become more common for health purposes.

Another critical consideration with passive systems deployed on ubiquitous devices is the need to minimize false positives. Generally, it is not wise to use these systems for asymptomatic screening of healthy populations given the dangers of excessive false-positives. Using these systems to monitor specific patient populations at risk for certain physiologic perturbations that are clinically meaningful are more likely to be useful to the patient and care teams generally. Toward this end, following identification of a given biomarker or aberrant trend, effective uses of these systems will likely require a level of interactivity (via screen or voice) to collect further information, such as pertinent positives and negatives, before consequential actions or referrals are executed. Additionally, as these computing systems mature as tools for research, they will require a research platform that can enable vetted, high-quality studies at scale, similar to Apple's ResearchKit, Sage Bionetwork's Bridge platform and MyDataHelps from Care Evolution. Such research is essential to demonstrate the health utility of these platforms, which will require actual clinical evidence to gain trust from patients, care teams and health systems. Finally, when used for health purposes, it is essential these devices do not exacerbate health disparities, for example by being differentially accessible to certain populations.

**Conclusion**

In summary, smart speakers represent a new, ubiquitous computing platform within our home environments, which holds considerable untapped potential to improve human health, at low cost and, if done thoughtfully, in ways that foster high compliance and preserve privacy. The primary health benefits are likely to be observed with enhanced chronic disease management, early detection of unwitnessed emergencies and indolent neurodegenerative processes, and enhancements of the patient/provider experience in clinic and inpatient environments. Realizing this unrealized potential will require smart speaker manufacturers to open their platforms to developers as they have with smartphones, develop an ecosystem specifically for medically-oriented applications and research, and enable and relentlessly prioritize privacy-preserving functionality.

**Funding:** JES is supported by NSF (1914873149 and 181255) and NIH (K23DA046686, R44DA050339). **Competing interests:** JES holds an equity stake in Sound Life Sciences, Inc. (Seattle, WA).